\documentclass[preprint, prX]{revtex4}

\usepackage{amsmath}    
\usepackage{graphicx}   
\usepackage{verbatim}   
\usepackage{color}      
\usepackage{subfigure}  
\usepackage{hyperref}   
\usepackage{amssymb}    
\usepackage{epsfig}
\usepackage{graphics,graphicx}
\usepackage{setspace}
\usepackage{url}
\usepackage{algorithm,algorithmic}
\usepackage{MnSymbol}

\begin{document}

\begin{abstract}
A manually time-measuring tool in mass sporting competitions cannot be imagined nowadays because many modern disciplines, such as
IronMan, take a long time and, therefore, demand additional reliability. Moreover, automatic timing devices, based on RFID technology,
have become cheaper. However, these devices cannot operate stand-alone because they need a computer measuring system that is capable of
processing the incoming events, encoding the results, assigning them to the correct competitor, sorting the results according to the
achieved times, and then providing a printout of the results. In this article, the domain-specific language EasyTime is presented, which enables the controlling of an agent by writing the events in a database. In particular, we are focused on the implementation of EasyTime with a LISA tool that enables the automatic construction of compilers from language specifications using Attribute Grammars. By using of EasyTime, we can also decrease the number of measuring devices. Furthermore, EasyTime is universal and can be applied to many different
sporting competitions in practice.

\textit{To cite paper as follows: I.~Jr. Fister, M.~Mernik, I.~Fister, and D.~Hrn\v{c}i\v{c},
Implementation of the domain-specific language easytime using a
{LISA} compiler generator. In {\em Proceedings of the Federated Conference on Computer Science
  and Information Systems}, pages 809--816, Szczecin, Poland, 2011.
}

\end{abstract}

\title{Implementation of the Domain-Specific Language EasyTime using a LISA Compiler Generator}

\author{Iztok Fister Jr.}
\altaffiliation{University of Maribor, Faculty of electrical engineering and computer science
Smetanova 17, 2000 Maribor}
\email{iztok.fister@guest.arnes.si}

\author{Marjan Mernik}
\altaffiliation{University of Maribor, Faculty of electrical engineering and computer science
Smetanova 17, 2000 Maribor}
\email{marjan.mernik@uni-mb.si}

\author{Iztok Fister}
\altaffiliation{University of Maribor, Faculty of electrical engineering and computer science
Smetanova 17, 2000 Maribor}
\email{iztok.fister@uni-mb.si}

\author{Dejan Hrn\v{c}i\v{c}}
\altaffiliation{University of Maribor, Faculty of electrical engineering and computer science
Smetanova 17, 2000 Maribor}
\email{dejan.hrncic@uni-mb.si}

\maketitle

\section{Introduction}
In the past, timekeepers measured the time manually. The time from a timer was assigned to competitors based on their starting number and these competitors were then ordered according to their achieved results and category. Later, the manual timers were replaced by the timers with an automatic time register that was capable of capturing and printing out registered times. However, an assigning the time to a competitor based on their starting number was still done manually. This work could be avoided by using the electronic measuring technology which, in addition to registering the time, also enabled the registering of the competitors' starting number. An expansion of RFID (Radio Frequency Identification) technology has helped this measuring technology become less expensive (\cite{web:ChampionChip2010,web:RFID2010}) and accessible to a wider range of users (e.g., sport clubs, organizers of sporting competitions). Moreover, they were able to compete with time-measuring monopolies at smaller competitions.

In addition to measuring technology, a flexible computer system is also needed to monitor the results. The proposed computer system enables the monitoring of different sporting competitions with a various number of measuring devices and measuring points, the online recording of events, the writing of results, as well as efficiency and security. The measuring device is dedicated to the registration of events and is triggered either automatically, when the competitor crosses the measuring point that acts as an electromagnetic antenna fields with an appropriate RFID tag, or manually, when an operator presses the suitable button on a personal computer that acts as a timer. The control point is the place where the organizers want to monitor results. Until now, each control point required its own measuring device. However, modern electronic measuring devices now allow for the handling of multiple control points simultaneously. Moreover, each registered event can have a different meaning, depending on the situation in which it is generated. Therefore, the event is handled by the measuring system according to the rules that are valid for the control point. As a result, the number of control points (and measuring devices) can be reduced with more complex measurements. Fortunately, the rules controlling events can be described easily with the use of a domain-specific language (DSL) \cite{Mernik:2005}. With this DSL, the measurements of different sporting competitions can be accomplished with the easy pre-configuration of rules.

A DSL is suited to an application domain and has certain advantages over general purpose languages (GPL) in a specific domain
\cite{Kosar:2010,Mernik:2005}. The GPL is dedicated to writing software in a wider range of application domains. With these languages
general problems are usually solved. However, to change the behavior of a program written in a GPL, a programmer is necessary. On the
other hand, the advantages of DSL are reflected in its greater expressive power and hence increased productivity, ease of use (even for domain experts that are not programmers), and easier verification and optimization \cite{Mernik:2005}. In this article, a DSL called EasyTime and it's implementation is presented. EasyTime is intended to control the agents that are responsible for recording events from the measuring devices into a database. Therefore, the agents are crucial elements of the proposed measuring system. Finally,
EasyTime was successfully employed in practice as well. For instance, it measured times in a World championship for the ultra double
triathlon in 2009~\cite{Fister:2011} and a National Championship in the time trials for bicycle in 2010~\cite{Fister:2011}.

The structure of the rest of the article is as follows; In the second section, the problems that are accompanied with time-measuring at sporting competitions are illustrated. We focus primarily on triathlon competitions, because they contain three disciplines that
need to be measured and also because of their long duration. The design of DSL EasyTime is briefly shown in section three. In the fourth section, the implementation of the EasyTime compiler is described, while in the fifth section the execution of the program written in
EasyTime is explained. Finally, the article is concluded with a short analysis of the work performed and a look at future work.

\section{Measuring Time in Sporting Competitions}
In practice, the measuring time in sporting competitions can be performed manually (classically or with a computer timer) or automatically (with a measuring device). The computer timer is a program that usually runs on a workstation (personal computer) and measures in real time. Thereby, a processor tact is exploited. The processor tact is the velocity with which the processor's instructions are interpreted. A computer timer enables the recording of events that are generated by the competitor crossing the measure points (MP) similar to the measuring device. In that case, however, the event is triggered by an operator pressing the appropriate button on the computer. The operator generates events in the form of $\langle\#,MP,TIME\rangle$, where $\#$ denotes the starting number of a competitor, $MP$ is the measuring point and $TIME$ is the number of seconds since 1.1.1970 at 0:0:0 (timestamp). One computer timer represents one measuring point.

Today, the measuring device is usually based on RFID (Radio Frequency Identification) technology \cite{Finkenzeller:2010}, where an
identification is performed with electromagnetic waves in the range of radio frequencies and consists of the following elements:

\begin{itemize}
  \item readers of RFID tags,
  \item primary memory,
  \item LCD monitor,
  \item numeric keyboard, and
  \item antenna fields.
\end{itemize}

More antenna fields can be connected on the measuring device. One antenna field represents one measuring point. Each competitor
generates an event by crossing the antenna field with passive RFID tags that include an identification number. This number is unique and differs from the starting number of the competitor. The event from the measuring device is represented in the form of
$\langle\#,RFID,MP,TIME\rangle$, where the identification number of the RFID tag is added to the previously mentioned triplet.

The measuring devices and workstations running the computer timer can be connected to the local area network. Communication with devices is performed by a monitoring program, i.e. an agent, that runs on the database server. The agent communicates with the measuring device
via the TCP/IP sockets and appropriate protocol. Usually, the measuring devices support a protocol $Telnet$ that is character-stream
oriented and, therefore, easy to implement. The agent employs the file transfer protocol to communicate with the computer timer.

\subsection{Example: Time Measuring Times at Triathlons}

Special conditions apply for triathlon competitions, where one competition consists of three disciplines. In this article,
therefore, we will devote the most of our attention to this problem.

The triathlon competition was first held in the USA in the year 1975. Today, the competition has become an Olympic discipline as well.
The triathlon competition is performed as follows: first, the athletes swim, then they ride a bike and finally they run. In practice,
all these activities are performed continuously. However, the transition times, i.e. the time that elapses when the competitor shifts
from swimming to bicycling and from bicycling to running, are added to the summary result. There are various types of triathlon
competitions that differ according to the length of various courses. To make things easier, the organizers will often employ the round
courses (laps) of shorter lengths instead of one long course. Therefore, the difficulty of measuring time is increased because the time
for each lap needs to be measured.

\begin{figure*}[htb]  
\vspace{-5mm}
    \begin{center}
        \includegraphics [scale=1.0]{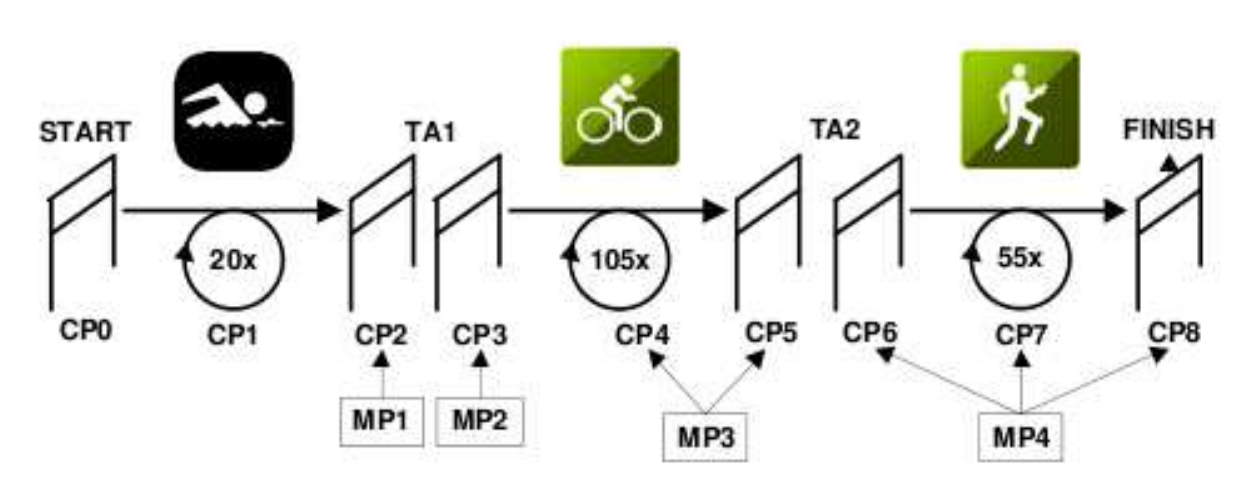}  %
        \caption{Definition of control points in the triathlon.}
        \label{pic:slika_1}
    \end{center}
\vspace{-5mm}
\end{figure*}

Measuring time in triathlon competitions can be divided into nine control points (Fig.~\ref{pic:slika_1}). The control point (CP) is a
location on the triathlon course, where the organizers need to check the measured time. This can be intermediate or final. As can be
seen in Fig.~\ref{pic:slika_1}, when dealing with a double triathlon there are 7.6 km of swimming, 360 km of bicycling and 84 km of
running, while the swimming course of 380 meters consists of 20 laps, the bicycling course of 3.4 kilometers contains 105 laps and the
running course of 1.5 kilometers has 55 laps.

Therefore, the final result of each competitor in a triathlon competition (CP8) consists of five final results: the swiming time SWIM
(CP2), the time for the first transition TA1 (CP3), the time spent bicycling BIKE (CP5), the time for the second transition TA2 (CP6),
the time spent running RUN (CP8), and three intermediate results: the intermediate time for swimming (CP1), the intermediate time for
bicycling (CP4) and the intermediate time for running (CP7). However, the current time INTER\_x and the number of remaining laps LAPS\_x are measured by the intermediate results, where $x=\{1,2,3\}$ denotes the appropriate discipline (1=SWIM, 2=BIKE and 3=RUN).

Suppose a measuring device with two measuring places (MP3 and MP4) is available to measure a triathlon competition as illustrated in
Fig.~\ref{pic:slika_1} and the competition is performed in one location. In that case, the last crossing over the MP3 denotes the time
of CP5, the first crossing over the MP4 denotes the time CP6 and the last crossing over the MP4 is the final time (CP8). The measuring
places MP1 and MP2 are conducted manually by the computer timers. The number of control points, thereby, can be reduced by three if the
measuring system and the appropriate setting of control points are employed. In fact, the 162 events pro competitor (87.6\% of all) can
be measured by this device. Moreover, because the technology for measuring swimmers in lakes and seas is very expensive and, therefore,
usually recorded by referees manually, this real-world competition would be covered by measuring the 98\% of all events.

In order to achieve this goal, however, the DSL EasyTime was developed and employed in practice by conducting measurements at the World Championship in Double Triathlon in 2009. Note that measurements were realized according to Fig.~\ref{pic:slika_1}. In the next sections, a design, an implementation, and an operation of EasyTime is presented.


\section{The Design of the Domain-Specific Language EasyTime}

Typically, the development of a DSL consists of the following phases:
\begin{itemize}
  \item a domain analysis,
  \item a definition of an abstract syntax,
  \item a definition of a concrete syntax,
  \item a definition of a formal semantics, and
  \item an implementation of the DSL.
\end{itemize}
Domain analysis provides an analysis of the application domain, i.e. measuring time in sporting competitions. The results of this
analysis define the concepts of EasyTime that are typically represented in a feature diagram~\cite{Deursen:2002}. The feature diagram
describes also dependencies between the concepts of the DSL. Thus, each concept can be broken down into features and sub-features. In
the case of EasyTime, the concept $race$ consists of the sub-features: $events$ (e.g., $swimming$, $bicycling$ and $running$), $control\ points$, $measuring\ time$, $transition\ area$, and $agents$. Each $control\ point$ is described by its $starting$ and $finish$ line and at least one $lap$. In addition, the feature $transition\ area$ can be introduced as the difference between the finish and start times. Both $updating\ time$ and $decrementing\ laps$ are sub-features of $measuring\ time$. However, an $agent$ for the processing of events received from the measuring device is needed. It can acts either $automatically$ or $manually$.

Domain analysis identifies several concepts in the application domain that needs to be mapped into EasyTime syntax and
semantics~\cite{Mernik:2005}. At first, the abstract syntax is defined (context-free grammar). Each concept obtained from the domain
analysis is mapped to a non-terminal in the context-free grammar; additionally, some new non-terminal and terminal symbols are defined. The translations of the EasyTime domain concepts to non-terminals are presented in Table~\ref{tab:tab1}, while an abstract syntax is presented in Table~\ref{tab:X}. Interestingly, a description of agents and measuring places cannot be found in other DSLs or GPLs. While attribute declaration is similar to variable declaration in many other programming languages there is the distinction that variables are actually database attributes allocated for every competitor. Some statements, such as assignment, conditional statement, and compound statement can be found in many other programming languages, while decrement attributes and update attributes are domain-specific constructs.

\begin{table*}[htb]        
\caption{Translation of the application domain concepts to a context-free grammar}
\label{tab:tab1}
\vspace{-5mm}
\small
\begin{center}
\scalebox{0.7} {
\begin{tabular}{ l  l  l  l }
\hline
Application domain concepts &  Non-terminal & Formal semantics & Description \\
\hline
Race & P & $\mathcal{CP}$ & Description of agents; control points; measuring \\
 & & & places. \\
\hline
Events (swimming, cycling, & none & none & Measuring time is independent from the type of an \\
running) & & & event. However, good attribute's identifier in control \\
 & & & points description will resemble the type of an event. \\
\hline
Transition area times & none & none & Can be computed as difference between events final \\
 & & & and starting times. \\
\hline
Control points (start, number & D & $\mathcal{D}$ & Description of attributes where start and finish time \\
of laps, finish) & & & will be stored as well as remaining laps. \\
\hline
Measuring places (update time, & M & $\mathcal{CM}$ & Measuring place id; agent id, which will control this \\
decrement lap) & & & measuring place; specific actions which will be per- \\
 & & & formed at this measuring place (e.g., decrement lap). \\
\hline
Agents (automatic, manual) & A & $\mathcal{A}$ & Agent id; agent type (automatic, manual); agent sour- \\
 & & & ce (file, ip). \\
\hline
\end{tabular}
}
\end{center}
\normalsize
\vspace{-5mm}
\end{table*}

\begin{table}[htb]           
\caption{The abstract syntax of EasyTime}
\label{tab:X}
\vspace{-5mm}
\begin{center}
\begin{tabular}{ | l  l  l | }
\hline
  $P \in $ \textbf{Pgm} & &  $A \in $ \textbf{Adec} \\
  $D \in $ \textbf{Dec} & &  $M \in $ \textbf{MeasPlace} \\
  $S \in $ \textbf{Stm} & &  $b \in $ \textbf{Bexp} \\
  $a \in $ \textbf{Aexp} & &  $n \in $ \textbf{Num} \\
  $x \in $ \textbf{Var} & &  $file \in $ \textbf{FileSpec} \\
  $ip \in $ \textbf{IpAddress} & &  \\
  & & \\
  $P$ & ::= & $A\ D\ M$ \\
  $A$ & ::= & $n$ \textbf{manual} $file$ \textbar $\ n$ \textbf{auto} $ip$ \textbar $\ A_{1};A_{2}$ \\
  $D$ & ::= & \textbf{var} $x := a$ \textbar $\ D_{1};D_{2}$ \\
  $M$ & ::= & \textbf{mp}[$n_{1}$] $\rightarrow$ \textbf{agnt}[$n_{2}]\ S$ \textbar $\ M_{1};M_{2}$ \\
  $S$ & ::= & \textbf{dec} $x$ \textbar \ \textbf{upd} $x$ \textbar $\ x := a$ \textbar $\ (b) \rightarrow S$ \textbar $\ S_{1};S_{2}$
  \\
  $b$ & ::= & \textbf{true} \textbar \ \textbf{false} \textbar $\ a_{1} = a_{2}$ \textbar $\ a_{1} != a_{2}$ \\
  $a$ & ::= & $n$ \textbar $\ x$ \\
\hline
\end{tabular}
\end{center}
\vspace{-5mm}
\end{table}

In the formal semantics phase, a meaning of the EasyTime language constructs is prescribed.
Each language construct, belonging to the syntax domain, is mapped into an appropriate semantic domain (Table~\ref{tab:Y}) by semantic
functions $\mathcal{CP}$, $\mathcal{CM}$, $\mathcal{CS}$, $\mathcal{A}$ (Table~\ref{tab:Z}). In addition, semantic functions
$\mathcal{A}$ and $\mathcal{CM}$ are illustrated by Table~\ref{tab:tab6}.

\begin{table}[htb]        
\caption{Semantic domains}
\label{tab:Y}
\vspace{-5mm}
\begin{center}
\begin{tabular}{ | l  l | }
\hline
  \textbf{Integer}=$\{\ldots -3,-2,-1,0,1,2,3 \ldots\}$ & $n \in$ \textbf{Integer} \\
  \textbf{Truth-Value}=$\{true,false\}$ &  \\
  \textbf{State}=\textbf{Var}$\rightarrow$\textbf{Integer} & $s \in$ \textbf{State} \\
  \textbf{AType}=$\{manual,auto\}$ &  \\
  \textbf{Agents}=\textbf{Integer}$\rightarrow$\textbf{AType}$\ \times\ (FileSpec\ \cup\ IpAddress)$ & $ag \in$\textbf{Agents} \\
  \textbf{Runners}=$(Id \times RFID \times LastName \times FirstName)^{*}$ & $r \in $ \textbf{Runners}\\
  \textbf{DataBase}=$(Id \times Var_{1} \times Var_{2} \times \ldots \times Var_{n})^{*}$ & $db \in $ \textbf{DataBase}\\
  \textbf{Code}=\textbf{String} & $c \in $ \textbf{Code} \\
\hline
\end{tabular}
\end{center}
\vspace{-5mm}
\end{table}

\begin{table}[htb]        
\caption{Translation of the syntax domain to semantic domains by semantic functions}
\label{tab:Z}
\vspace{-5mm}
\begin{center}
\begin{tabular}{ | l  l  l | }
\hline
 Syntax Domain & Semantic Function & Semantic Domain \\
\hline
 \textbf{Pgm} & $\mathcal{CP}$ & \textbf{Code} $\times$ \textbf{Integer} $\times$ \textbf{Database} \\
 \textbf{MeasPlace} & $\mathcal{CM}$ & \textbf{Code} $\times$ \textbf{Integer} \\
 \textbf{Stm} & $\mathcal{CS}$ & \textbf{Code} \\
 \textbf{Adecs} & $\mathcal{A}$ & \textbf{Agents} \\
\hline
\end{tabular}
\end{center}
\vspace{-5mm}
\end{table}

\begin{table}[htb]
\caption{Translation of agents and measuring places}
\label{tab:tab6}
\vspace{-5mm}
\begin{center}
\begin{tabular}{ | l  l  l | }
\hline
  $\mathcal{A}$:\textbf{Adec} $\rightarrow$ \textbf{Agents} & $\rightarrow$ & \textbf{Agents} \\
  $\mathcal{A} \lsem n$ \textbf{manual} $ file\rsem ag$ & = & $ag [ n \rightarrow (manual, file) ]$ \\
  $\mathcal{A} \lsem n$ \textbf{auto} $ ip\rsem ag$ & = & $ag [ n \rightarrow (auto, ip) ]$ \\
  $\mathcal{A} \lsem A_{1};A_{2}\rsem ag$ & = & $\mathcal{A} \lsem A_{2} \rsem (\mathcal{A} \lsem A_{1} \rsem ag)$ \\
  & & \\
  $\mathcal{CM}$:\textbf{MeasPlace} $\rightarrow$ \textbf{Agents}&$\rightarrow$&\textbf{Code} $\times$ \textbf{Integer} \\
  $\mathcal{CM} \lsem \textbf{mp}[n_{1}] \rightarrow \textbf{agnt}[n_{2}] S \rsem ag$&=&(WAIT $i:\mathcal{CS} \lsem S \rsem (ag, n_{2}),
  n_{1} )$  \\
  $\mathcal{CM} \lsem M_{1}; M_{2} \rsem ag$&=&$\mathcal{CM} \lsem M_{1} \rsem ag: \mathcal{CM} \lsem M_{2} \rsem ag$ \\
\hline
\end{tabular}
\end{center}
\vspace{-5mm}
\end{table}

The sample program written in EasyTime that covers the measuring time in the double ultra triathlon as illustrated by
Fig.~\ref{pic:slika_1} is presented by Algorithm~\ref{alg:prog}.

\begin{algorithm}[htb]
\caption{EasyTime program for measuring time in a triathlon competition as illustrated in Fig.~\ref{pic:slika_1}}
\label{alg:prog}
\small
\begin{algorithmic}[1]
\STATE 1 manual "abc.res";
\STATE 2 auto 192.168.225.100;
\STATE
\STATE var ROUND1 := 20;
\STATE var INTER1 := 0;
\STATE var SWIM := 0;
\STATE var TRANS1 :=0;
\STATE var ROUND2 := 105;
\STATE var INTER2 :=0;
\STATE var BIKE := 0;
\STATE var TRANS2 :=0;
\STATE var ROUND3 := 55;
\STATE var INTER3 := 0;
\STATE var RUN := 0;
\STATE
\STATE mp[1] $\rightarrow$ agnt[1] \{
\STATE \ \ (true) $\rightarrow$ upd SWIM;
\STATE \ \ (true) $\rightarrow$ dec ROUND1;
\STATE \}
\STATE mp[2] $\rightarrow$ agnt[1] \{
\STATE \ \ (true) $\rightarrow$  upd TRANS1;
\STATE \}
\STATE mp[3] $\rightarrow$  agnt[2] \{
\STATE \ \ (true) $\rightarrow$  upd INTER2;
\STATE \ \ (true) $\rightarrow$  dec ROUND2;
\STATE \ \ (ROUND2 == 0) $\rightarrow$  upd BIKE;
\STATE \}
\STATE mp[4] $\rightarrow$  agnt[2] \{
\STATE \ \ (true) $\rightarrow$  upd INTER3;
\STATE \ \ (ROUND3 == 55) $\rightarrow$  upd TRANS2;
\STATE \ \ (true) $\rightarrow$  dec ROUND3;
\STATE \ \ (ROUND3 == 0) $\rightarrow$  upd RUN;
\STATE \}
\end{algorithmic}
\normalsize
\end{algorithm}

More details of EasyTime syntax and semantics are presented in~\cite{Fister:2011}. In this article, we are focused on the implementation
phase as presented in the next section.

\section{The Implementation of the Domain-Specific Language EasyTime}
\begin{figure*}[htb]      
    \begin{center}
        \includegraphics [scale=0.6] {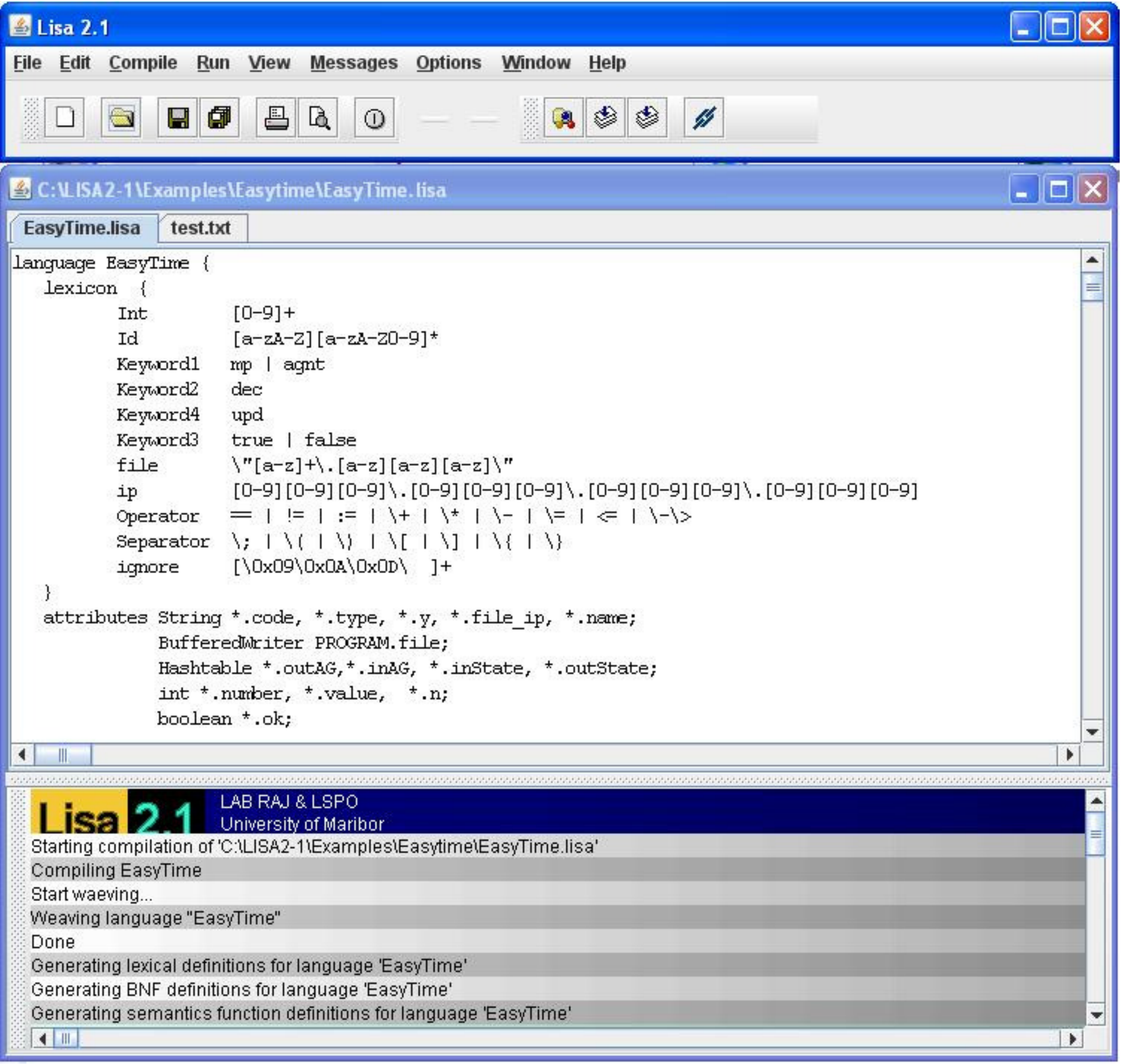}  
        \caption{LISA GUI.}
        \label{pic:LISA_GUI}
    \end{center}
\vspace{-5mm}
\end{figure*}

Our motivation was to automatize an implementation phase as much as possible. Therefore, we use a compiler generator that can convert a
formal description of a programming language into a compiler/interpreter for that language. Several recent compiler generators accept
descriptions in terms of attribute grammars or denotational semantics~\cite{Paulson:1982}. Although many compiler generators exist
today, we selected a LISA compiler-compiler that was developed at the University of Maribor in the late 1990s~\cite{Mernik:2002}. The LISA tool produces highly efficient source code for: scanner, parser, interpreter or compiler in Java. The lexical and syntactical parts of a language specification in LISA supports various well known formal methods, like regular expressions and BNF. LISA provides two kinds of user interfaces:
\begin{itemize}
  \item a graphic user interface (GUI) (Fig.~\ref{pic:LISA_GUI}), and
  \item a Web-Service user interface.
\end{itemize}

The main features of LISA are as follows:
\begin{itemize}
 \item since it is written in Java, LISA works on all Java platforms,
 \item a textual or a visual environment,
 \item an Integrated Development Environment (IDE), where users can specify, generate, compile and execute programs on the fly,
 \item visual presentations of different structures, such as finite-state-automata, BNF, a dependency graph, a syntax tree, etc.,
 \item modular and incremental language development~\cite{Mernik:2005a}.
\end{itemize}

LISA specifications are based on Attribute Grammar (AG)~\cite{Paakki:1995} that has been introduced by D.E. Knuth~\cite{Knuth:1968}.
The attribute grammar is a triple $AG=\langle G,A,R \rangle$, where $G$ denotes a context-free grammar, $A$ a finite set of attributes
and $R$ a finite set of semantic rules. In line with this, LISA specifications include:
\begin{itemize}
  \item lexical regular definitions,
  \item attribute definitions,
  \item semantic rules, and
  \item operations on semantic domains.
\end{itemize}

Lexical specifications for EasyTime in LISA (Figure \ref{pic:LISA_GUI}) are similar to those used in other compiler generators.
While LISA automatically infers whether an attribute is inherited or synthesized \cite{Knuth:1968}, the type of an attribute must
be specified (Figure \ref{pic:LISA_GUI}). For example, the attribute \textit{code} represents generated code using translation
functions, the attribute \textit{outAG} is the synthesized attribute and \textit{inAG} the inherited attribute representing agents (\textit{ag} from semantic specifications). The correspondence between attributes in LISA specifications and a semantic description of
EasyTime is shown in Table~\ref{tab:W}.

\begin{table}[htb]
\caption{Translating of semantic function to LISA specifications}
\label{tab:W}
\vspace{-5mm}
\begin{center}
\begin{tabular}{ l l l }
\hline
 Semantic Spec. & Semantic Domain & LISA Attributes \\
\hline
 \textit{c} & \textbf{Code} & String *.code \\
 \textit{ag} & \textbf{Agents} & Hashtable *.inAG, *.outAG \\
 \textit{s} & \textbf{State} & Hashtable *.inState, *.outState \\
 \textit{n} & \textbf{Integer}& *.n  \\
\hline
\end{tabular}
\end{center}
\vspace{-5mm}
\end{table}

Essentially, we are focused on LISA specifications of semantic rules, which consists of generalized syntax rules that also encapsulate
semantic rules. The semantic rules of EasyTime, as presented in Section III, were translated into the LISA specifications according to
Table~\ref{tab:tab2}.

\begin{table}[htb]
\caption{Translating of semantic functions to LISA specifications}
\label{tab:tab2}
\vspace{-5mm}
\begin{center}
\begin{tabular}{ l l }
\hline
 Semantic Function & LISA Specification \\
\hline
 $\mathcal{CP}$ & Start \\
 $\mathcal{CM}$ & Mes\_Places \\
 $\mathcal{CS}$ & Stmts \\
 $\mathcal{A}$ & Agents \\
\hline
\end{tabular}
\end{center}
\end{table}

Due to page limitations, only some part of mapping of EasyTime semantic specifications into LISA specifications are explained in this
paper.

During the conversion from the abstract syntax to the concrete syntax, the production in the abstract syntax $A_1;A_2$ denoting a sequence of agents, is translated to the following production in the concrete syntax:
\footnotesize
\begin{equation*}
\mathit{AGENTS}\ ::=\ \mathit{AGENTS}\ \mathit{AGENT\ |\ \varepsilon}.
\end{equation*}
\normalsize
The semantic function $\mathcal{A} \lsem A_{1};A_{2}\rsem ag$ = $\mathcal{A} \lsem A_{2} \rsem (\mathcal{A} \lsem A_{1} \rsem ag)$
constructs $ag \in Agents$, which is a function from an integer, denoting an agent, into an agent's type (manual or auto) and an agent's ip or agent's file. This function is described in LISA for non epsilon cases as:
\footnotesize
\begin{equation*}
    AGENTS[1].inAG\ =\ AGENTS[0].inAG;
\end{equation*}
\begin{equation*}
    AGENTS[0].outAG\ =\ insert(AGENTS[1].outAG,new\\
\end{equation*}
\begin{equation*}
    Agent(AGENT.number,\ AGENT.type,\ AGENT.file\_ip));\\
\end{equation*}
\normalsize
and for epsilon cases as:
\footnotesize
\begin{equation*}
    AGENTS.outAG\ =\ AGENTS.inAG;.
\end{equation*}
\normalsize
The net effect is that we are constructing a list, more precisely a hash table, of agents where we are recording the agent's number ($AGENT.number$), the agents's type ($AGENT.type$), and the agent's ip or file ($AGENT.file\_ip$). Those attributes are defined
in the productions:
\footnotesize
\begin{equation*}
    AGENT ::= \#Int\ auto\ \#ip;\ |\ \#Int\ manual\ \#file;
\end{equation*}
\normalsize
(Algorithm \ref{alg:agent_lisa}) and implements semantic functions:
\footnotesize
\begin{equation*}
\mathcal{A} \lsem n\ \textbf{manual}\ file\rsem ag\ =\ ag [ n \rightarrow (manual, file) ]
\end{equation*}
\normalsize
and
\footnotesize
\begin{equation*}
\mathcal{A} \lsem n\ \textbf{auto}\ ip\rsem ag\ =\ ag [ n \rightarrow (auto, ip) ].
\end{equation*}
\normalsize

\begin{algorithm}[tbh]
\caption{Translation of Agents into LISA specifications}
\label{alg:agent_lisa}
\footnotesize
\begin{algorithmic}[1]
\STATE rule Agents \{
\STATE \ \ \ \ AGENTS ::= AGENTS  AGENT compute \{
\STATE \ \ \ \ \ \ \ \ AGENTS[1].inAG = AGENTS[0].inAG;
\STATE \ \ \ \ \ \ \ \ AGENTS[0].outAG = insert(AGENTS[1].outAG,
\STATE \ \ \ \ \ \ \ \ new Agent(AGENT.number, AGENT.type, AGENT.file\_ip));
\STATE \ \ \ \ \}
\STATE \ \ \ \ $|$ epsilon compute \{
\STATE \ \ \ \ \ \ \ \ AGENTS.outAG = AGENTS.inAG;
\STATE \ \ \ \ \};
\STATE \}
\STATE rule AGENT \{
\STATE \ \ \ \ AGENT ::= \#Int manual \#file \; compute \{
\STATE \ \ \ \ \ \ \ \ AGENT.number = Integer.valueOf(\#Int[0].value()).intValue();
\STATE \ \ \ \ \ \ \ \ AGENT.type = "manual";
\STATE \ \ \ \ \ \ \ \ AGENT.file\_ip = \#file.value();
\STATE \ \ \ \ \};
\STATE \ \ \ \ AGENT ::= \#Int auto \#ip \; compute \{
\STATE \ \ \ \ \ \ \ \ AGENT.number = Integer.valueOf(\#Int[0].value()).intValue();
\STATE \ \ \ \ \ \ \ \ AGENT.type = "auto";
\STATE \ \ \ \ \ \ \ \ AGENT.file\_ip = \#ip.value();
\STATE \ \ \ \ \};
\STATE \}
\end{algorithmic}
\normalsize
\end{algorithm}

During the conversion from the abstract syntax to the concrete syntax, the production in the abstract syntax $M_1;M_2$
denoting a sequence of measuring places is translated to the following production
in the concrete syntax:
\footnotesize
\begin{equation*}
\mathit{MES\_PLACES}\ ::=\ \mathit{MES\_PLACE}\ \mathit{~MES\_PLACES\ |\ MES\_PLACE}.
\end{equation*}
\normalsize
The translation function:
\footnotesize
\begin{equation*}
\mathcal{CM} \lsem M_{1} \rsem ag\ :\ \mathcal{CM} \lsem M_{2} \rsem ag
\end{equation*}
\normalsize
translates into code the first construct $M_1$ before the translation of the second construct $M_2$ is performed. This function is
described in LISA as:
\footnotesize
\begin{equation*}
\mathit{MES\_PLACES}[0].code = \mathit{MES\_PLACE.code} +
\end{equation*}
\begin{equation*}
''\backslash n'' + \mathit{MES\_PLACES[1].code};
\end{equation*}
\normalsize
with the following meaning: The code for the first construct $\mathit{MES\_PLACE}$ is simply concatenated with the code from the second construct $MES\_PLACES[1]$. While the abstract syntax for the definition of the measuring place:
\footnotesize
\begin{equation*}
\textbf{mp}[n_{1}] \rightarrow \textbf{agnt}[n_{2}] S
\end{equation*}
\normalsize
is translated to the following production in the concrete syntax:
\footnotesize
\begin{equation*}
\mathit{MES\_PLACE} ::= mp ~[ ~\#Int~ ] ~->~  agnt ~[ ~\#Int~ ] ~\{ ~STMTS ~\}.
\end{equation*}
\normalsize
The translation function:
\footnotesize
\begin{equation*}
(WAIT\ i:\mathcal{CS} \lsem S \rsem (ag, n_{2}), n_{1} )
\end{equation*}
\normalsize
is described in LISA as:
\footnotesize
\begin{equation*}
\mathit{MES\_PLACE.code}\ =\ ''WAIT\ i\ '' + STMTS.code +
\end{equation*}
\begin{equation*}
'','' + \#Int[0].value() + '')''; .
\end{equation*}
\normalsize
Note, that in the implementation of this semantic function (Algorithm \ref{alg:mp_lisa}) many other attributes need to be defined. For
example, a list of agents need to be propagated into statements ($STMTS.inAG$), as well as a list of database attributes ($STMTS.inState$).

\begin{algorithm}[tbh]
\caption{Translation of MES\_PLACE into LISA specifications}
\label{alg:mp_lisa}
\footnotesize
\begin{algorithmic}[1]
\STATE rule Mes\_places \{
\STATE \ \ \ \ MES\_PLACES ::= MES\_PLACE MES\_PLACES compute \{
\STATE \ \ \ \ \ \ \ \ MES\_PLACE.inAG = MES\_PLACES[0].inAG;
\STATE \ \ \ \ \ \ \ \ MES\_PLACES[1].inAG = MES\_PLACES[0].inAG;
\STATE \ \ \ \ \ \ \ \ MES\_PLACE.inState = MES\_PLACES[0].inState;
\STATE \ \ \ \ \ \ \ \ MES\_PLACES[1].inState = MES\_PLACES[0].inState;
\STATE \ \ \ \ \ \ \ \ MES\_PLACES[0].ok = MES\_PLACE.ok \&\& MES\_PLACES[1].ok;
\STATE \ \ \ \ \ \ \ \ MES\_PLACES[0].code = MES\_PLACE.code + "$\backslash$n" + \\\hspace{3.7cm} MES\_PLACES[1].code;
\STATE \ \ \ \ \};
\STATE MES\_PLACES ::=  MES\_PLACE compute \{
\STATE \ \ \ \ \ \ \ \ MES\_PLACE.inAG = MES\_PLACES.inAG;
\STATE \ \ \ \ \ \ \ \ MES\_PLACE.inState = MES\_PLACES.inState;
\STATE \ \ \ \ \ \ \ \ MES\_PLACES.ok = MES\_PLACE.ok;
\STATE \ \ \ \ \ \ \ \ MES\_PLACES.code = MES\_PLACE.code;
\STATE \ \ \ \ \};
\STATE \}
\STATE rule MES\_PLACE \{
\STATE \ \ \ \ MES\_PLACE ::= mp $\backslash[$ \#$\mathit{Int}$ $\backslash]$ $\backslash-\backslash>$ \\\hspace{2.7cm} $\mathit{agnt}$
$\backslash[$ \#$\mathit{Int}$ $\backslash]$ $\backslash\{$ STMTS $\backslash\}$ compute \{
\STATE \ \ \ \ \ \ \ \ STMTS.inAG = MES\_PLACE.inAG;
\STATE \ \ \ \ \ \ \ \ STMTS.inState = MES\_PLACE.inState;
\STATE \ \ \ \ \ \ \ \ STMTS.n = Integer.valueOf(\#Int[1].value()).intValue();
\STATE \ \ \ \ \ \ \ \ MES\_PLACE.ok = STMTS.ok;
\STATE \ \ \ \ \ \ \ \ MES\_PLACE.code = "(WAIT i " + STMTS.code + ", " + \\\hspace{3.2cm} \#Int[0].value() + ")";
\STATE \ \ \ \ \};
\STATE \}
\end{algorithmic}
\normalsize
\end{algorithm}

Attributes that represent semantic information belong to various semantic domains (Figure \ref{pic:LISA_GUI}).
The attributes in LISA can be objects of classes specified in the library with already defined behavior (e.g., Hashtable) or can be objects of user-defined classes. For example, the previously mentioned semantic domain $Agents$, can be implemented as a hash table, where each element is an instance of the class Agent (Algorithm \ref{alg:m_ag_lisa}), where the agents's number, type and ip or file are stored. Moreover, various operations over semantic domain (e.g., insert into hash table - Algorithm \ref{alg:m_ins_lisa}) can be easily
implemented using object-oriented programming. In Algorithm \ref{alg:m_ins_lisa}, it first checks if the agent is already defined. If this condition is not met a new agent is put into hash table.


\begin{algorithm}[tbh]
\caption{LISA definition of the semantic domain Agents}
\label{alg:m_ag_lisa}
\footnotesize
\begin{algorithmic}[1]
\STATE method M\_Agent \{
\STATE \ \ class Agent \{
\STATE \ \ \ \ int number;
\STATE \ \ \ \ String type;
\STATE \ \ \ \ String file\_ip;
\STATE \ \ \ \ Agent ( int number, String type, String file\_ip) \{
\STATE \ \ \ \ \ \ this.number = number;
\STATE \ \ \ \ \ \ this.type = type;
\STATE \ \ \ \ \ \ this.file\_ip = file\_ip;
\STATE \ \ \ \ \}
\STATE \ \ \ \ public String toString() \{
\STATE \ \ \ \ \ \ return "(" + this.number + ", " + this.type + ", " + this.file\_ip + ")";
\STATE \ \ \ \ \}
\STATE \ \ \ \ public int getNumber() \{
\STATE \ \ \ \ \ \ return this.number;
\STATE \ \ \ \ \}
\STATE \ \ \ \ public String getType() \{
\STATE \ \ \ \ \ \ return this.type;
\STATE \ \ \ \ \}
\STATE \ \ \ \ public String getFile\_ip() \{
\STATE \ \ \ \ \ \ return this.file\_ip;
\STATE \ \ \ \ \}
\STATE \ \ \} $//$ Java class
\STATE \} $//$ Lisa method
\end{algorithmic}
\normalsize
\end{algorithm}

\begin{algorithm}[tbh]
\caption{Definition of the method Insert}
\label{alg:m_ins_lisa}
\footnotesize
\begin{algorithmic}[1]
\STATE method M\_Insert ( \{
\STATE \ \ import java.util.*;
\STATE \ \ Hashtable insert (Hashtable aAgents, Agent aAgent) \{
\STATE \ \ \ \ aAgents = (Hashtable)aAgents.clone();
\STATE \ \ \ \ Agent hAgent=(Agent)aAgents.get(aAgent.getNumber());
\STATE \ \ \ \ if (hAgent==null)
\STATE \ \ \ \ \ \ aAgents.put(aAgent.getNumber(), aAgent);
\STATE \ \ \ \ else
\STATE \ \ \ \ \ \ System.out.println("Agent" + aAgent.getNumber() + "is already defined");
\STATE \ \ \ \ return aAgents;
\STATE \ \ \} $//$ Java method
\STATE \} $//$ Lisa method
\end{algorithmic}
\normalsize
\end{algorithm}

\section{Operation}

Local organizers of sporting competitions were faced with two possibilities before the developing of EasyTime:
\begin{itemize}
  \item to rent a specialized company to measure time,
  \item to measure time manually.
\end{itemize}
The former possibility is expensive, while the latter can be very unreliable. However, the both objectives (i.e. inexpensiveness and reliability), can be fulfilled by EasyTime. On the other hand, producers of measuring devices usually deliver these units with software for collecting of events into a database. Then these events need to be post-processed (batch processed) to get the final results of competitors. Although this batch processing can be executed whenever the organizer desires each real-time application requests online processing. Fortunately, EasyTime enables both kinds of event processing.

In order to use the source program written in EasyTime by the measuring system, it needs to be compiled. Note that the code
generation \cite{Aho:1972} of a program in EasyTime is performed only if the parsing is finished successfully. Otherwise the compiler
prints out an error message and stops. For each measuring places individually, the code is generated by strictly following the rules, as defined in section III. An example of the generated code from the Algorithm~\ref{alg:prog} for controlling of the measurements,
as illustrated by Fig.~\ref{pic:slika_1}, is presented in Table~\ref{tab:tab10}. Note that the generated code is saved
into a database.

\begin{table}[htb]
\caption{Translated code for the EasyTime program in Algorithm~\ref{alg:prog}}
\label{tab:tab10}
\begin{center}
\small
\begin{tabular}{ | l | }
\hline

(WAIT i FETCH accessfile("abc.res") STORE SWIM \\
FETCH ROUND1 DEC STORE ROUND1, 1) \\ \\
(WAIT i FETCH accessfile("abc.res") STORE TRANS1, 2) \\ \\
(WAIT i FETCH connect(192.168.225.100) STORE INTER2 \\
FETCH ROUND2 DEC STORE ROUND2 \\
PUSH 0  FETCH ROUND2  EQ BRANCH( FETCH \\
connect(192.168.225.100) STORE BIKE, NOOP), 3) \\ \\
(WAIT i FETCH connect(192.168.225.100) STORE INTER3 \\
PUSH 55  FETCH ROUND3  EQ BRANCH( FETCH \\
connect(192.168.225.100) STORE TRANS2, NOOP) \\
FETCH ROUND3 DEC STORE ROUND3 \\
PUSH 0  FETCH ROUND3  EQ BRANCH( FETCH \\
connect(192.168.225.100) STORE RUN, NOOP), 4) \\

\hline
\end{tabular}
\normalsize
\end{center}
\end{table}

As a matter of fact, the generated code is dedicated to the control of an agent by writing the events received from the measuring
devices into the database. Typically, the program code is loaded from the database only once. That is, only an
interpretation of code could have any impact on the performance of a measuring system. Because this interpretation is not time consuming, it cannot degrade the performance of the system. On the other hand, the precision of measuring time is handled by the measuring device and it is not changed by the processing of events. In fact, the events can be processed as follows:
\begin{itemize}
  \item batch: manual mode of processing, and
  \item online: automatic mode of processing.
\end{itemize}
The agent reads and writes events that are collected in a text file when the first mode of processing is assumed. Typically, events
captured by a computer timer are processed in this mode. Here, the agent looks for the existence of the event text file that is
configured in the agent statement. If it exists, the batch processing is started. When the processing is finished, the text file is
archived and then deleted. The online processing is event oriented, i.e. each event that is generated by the measuring device is
processed in time.

\begin{figure}[htb]
\vspace{-5mm}
    \begin{center}
        \includegraphics [scale=0.75]{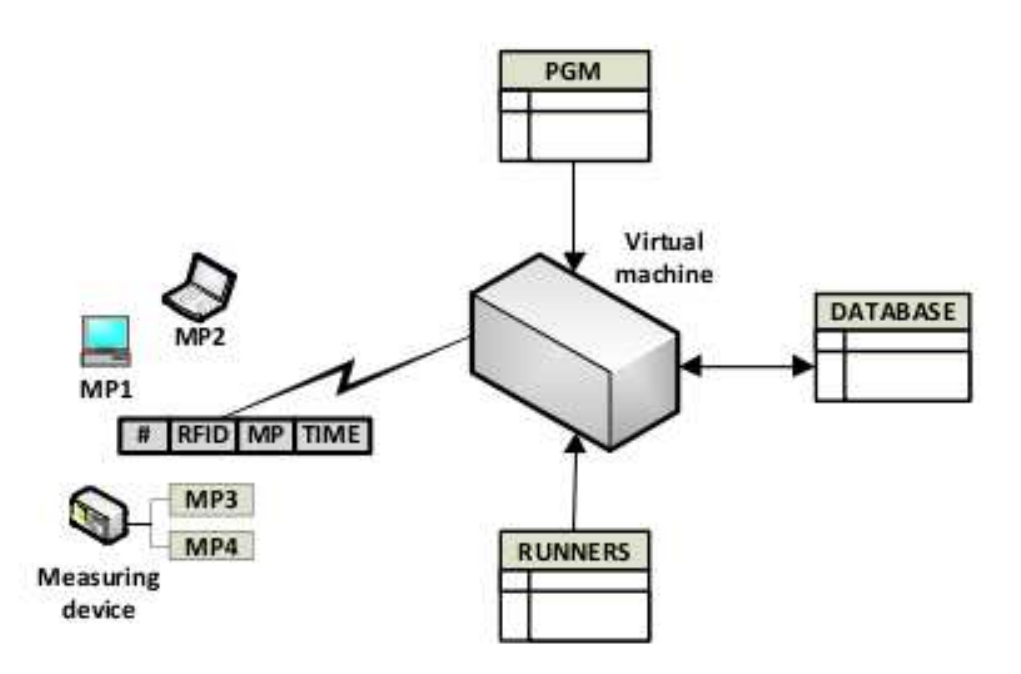}      %
        \caption{Executable environment of a program in EasyTime.}
        \label{pic:slika_3}
    \end{center}
\vspace{-5mm}
\end{figure}

In both modes of processing, the agent works with the program PGM, the runner table RUNNERS, and the results table DATABASE, as can be
seen in Fig.~\ref{pic:slika_3}. An initialization of the virtual machine is performed when the agent starts. The initialization consists of loading the program code from PGM. That is, the code is loaded only once. At the same time, the variables are initialized on starting values. A recording of events that are processed by the agent can be divided into the following phases:
\begin{itemize}
  \item Reconstruction of the event: the competitor is identified by a starting number ($\#$) or $RFID$ tag, $MP$ determines a
      virtual machine on which an interpretation of code will be run and the $TIME$ represents the timestamp of the event.
  \item Reading of results: the number ($\#$) or $RFID$ tag determines the competitor whose results are read from the table RUNNERS
      in the database.
  \item Mapping of the result: the read results are mapped into the data segment of the virtual machine that is identified by the
      $MP$. In addition, the program register is loaded with the timestamp $TIME$ of the event.
  \item Interpretation of code: the instruction counter is set to zero and the program loaded in the program segment of the virtual
      machine is started.
  \item Writing of results: after the interpretation of code, the results from the data segment are saved into the table DATABASE.
\end{itemize}

\section{Conclusion}
The flexibility of the measuring system is a crucial objective in the development of universal software for measuring time in sporting
competitions. Therefore, the domain-specific language EasyTime was formally designed, which enables the quick adaptation of a measuring
system to the new requests of different sporting competitions. Preparing the measuring system for a new sporting competition with EasyTime requires the following: changing a program's source code that controls the processing of an agent, compiling a source code and restarting the agent. Using EasyTime in the real-world had shown that when measuring times in a small sporting competitions, the organizers do not need to employ specialized and expensive companies any more. On the other hand, EasyTime can reduce the heavy configuration tasks of a measuring system for larger competitions as well. In this paper, we explained how the formal semantics of EasyTime are mapped into LISA specifications from which a compiler is automatically generated. Despite the fact that mapping is not difficult, it is not trivial either, as some additional rules must be defined for attribute propagation. Moreover, we need to take care of error reporting (eg., multiple definitions of agents). In future work, EasyTime could be replaced by the domain-specific modeling language (DSML) that could additionally simplify the programming of a measuring system.

\bibliographystyle{plain}
\bibliography{references}

\bigskip{\small \smallskip\noindent Updated 9 June 2012.}
\end{document}